\documentclass[11pt]{article}
\usepackage[a4paper]{geometry}
\usepackage{amsfonts, amsmath, amssymb, amsthm, graphicx, caption, authblk, multirow, makecell, framed, float, xcolor, enumitem, tikz, hyperref}
\theoremstyle{definition}

\theoremstyle{remark}

\newcommand*{\mybox}[1]{%
  \framebox{\raisebox{0cm}[0.5\baselineskip][0.05\baselineskip]{%
    \hbox to 0.10cm {\hss#1\hss}}}\hspace{0.05cm}}

\begin{document}
\title{The Landscape of Computing Symmetric $n$-Variable Functions with $2n$ Cards}
\author[1]{Suthee Ruangwises\thanks{\texttt{ruangwises@gmail.com}}}
\affil[1]{Department of Informatics, The University of Electro-Communications, Tokyo, Japan}
\date{}
\maketitle

\begin{abstract}
Secure multi-party computation using a physical deck of cards, often called card-based cryptography, has been extensively studied during the past decade. Card-based protocols to compute various Boolean functions have been developed. As each input bit is typically encoded by two cards, computing an $n$-variable Boolean function requires at least $2n$ cards. We are interested in optimal protocols that use exactly $2n$ cards. In particular, we focus on symmetric functions. In this paper, we formulate the problem of developing $2n$-card protocols to compute $n$-variable symmetric Boolean functions by classifying all such functions into several NPN-equivalence classes. We then summarize existing protocols that can compute some representative functions from these classes, and also solve some open problems in the cases $n=4$, 5, 6, and 7. In particular, we develop a protocol to compute a function $k$Mod3, which determines whether the sum of all inputs is congruent to $k$ modulo 3 ($k \in \{0,1,2\}$).

\textbf{Keywords:} card-based cryptography, secure multi-party computation, symmetric function
\end{abstract}

\section{Introduction}
Secure multi-party computation involves computing the output value of a particular function with inputs from different parties, while keeping the input values secret. Secure multi-party computation using a physical deck of cards, often called \textit{card-based cryptography}, has been a subject of research since the \textit{five-card trick} was introduced by den Boer \cite{denboer} in 1989. This area has gained interest from researchers and has been extensively studied during the past decade \cite{landscape}.

Typically, each input bit is encoded by an order of a black card \mybox{$\clubsuit$} and a red card \mybox{$\heartsuit$}; a bit 0 is represented by a \textit{commitment} \hbox{\mybox{$\clubsuit$}\mybox{$\heartsuit$}}, and bit 1 by a commitment \hbox{\mybox{$\heartsuit$}\mybox{$\clubsuit$}}. The five-card trick can compute a logical AND function of two input bits using five cards: two cards for a commitment of each bit, and one additional helping card. Later, Mizuki et al. \cite{mizuki12} showed that the two-variable AND function can be computed with only four cards, using no helping card.

Besides the AND function, protocols to compute various other Boolean functions have also been developed. As each input bit is encoded by two cards, computing an $n$-variable Boolean function requires at least $2n$ cards. A challenging work is to develop card-minimal protocols that use exactly $2n$ cards.

\section{Symmetric Boolean Functions}
We focus on \textit{symmetric} functions where each party is treated equally. A Boolean function $f:\{0,1\}^n \rightarrow \{0,1\}$ is called symmetric if $$f(x_1,x_2,...,x_n)=f(x_{\sigma(1)},x_{\sigma(2)},...,x_{\sigma(n)})$$ for any $x_1,x_2,...,x_n$ and any permutation $(\sigma(1),\sigma(2),...,\sigma(n))$ of $(1,2,...,n)$. Note that for a symmetric Boolean function $f$, the value of $f(x_1,x_2,...,x_n)$ only depends on the sum $\sum_{i=1}^n x_i$.

We denote an $n$-variable symmetric Boolean function by $S_X^n$ for some $X \subseteq \{0,1,...,n\}$. A function $S_X^n$ is defined by
$$S_X^n(x_1,x_2,...,x_n) = \begin{cases}
	1, &\text{ if } \sum_{i=1}^n x_i \in X; \\
  0, &\text{ otherwise.}
\end{cases}$$
For example, a function $x_1 \wedge x_2 \wedge ... \wedge x_n$ is denoted by $S_{\{n\}}^n$, and a function $x_1 \oplus x_2 \oplus ... \oplus x_n$ is denoted by $S_{\{1,3,5,...,2\lfloor\frac{n-1}{2}\rfloor+1\}}^n$.

In general, observe that if a function $f$ is computable by some number of cards, functions derived from $f$ by (1) negating variables, (2) permuting variables, and (3) negating the output, are also computable by the same number of cards. Hence, we can classify all $n$-variable symmetric Boolean functions into several classes called \textit{Negation-Permutation-Negation (NPN)-equivalence classes} \cite{class}, where all functions in the same class can be computed by essentially the same protocol, so it is sufficient to consider only one representative function from each class. Note that $S_X^n$ is always in the same class as $S_{\{n-x|x \in X\}}^n$ (negating all variables) and $S_{\{1,2,...,n\}-X}^n$ (negating $f$).

\section{Existing Protocols}
In 2015, Nishida et al. \cite{nishida} showed that any $n$-variable symmetric Boolean function can be computed with $2n+2$ cards, providing an upper bound for the number of required cards. However, we are mainly interested in $2n$-card protocols.

\subsection{Two Variables}
\begin{table}
	\centering
	\begin{tabular}{|c|c|c|c|c|c|}
		\hline
		\textbf{Function} & \textbf{Name} & \textbf{Protocol} & \textbf{\thead{Com-\\mitted?}} & \textbf{\thead{\#Shuf-\\fles}} & \textbf{\thead{Other Functions\\in the Same Class}} \\ \hline
		$S^2_{\emptyset}$ & Constant & \multicolumn{3}{c|}{trivial} & $S^2_{\{0,1,2\}}$ \\ \hline
		$S^2_{\{1\}}$ & XOR & Mizuki-Sone \cite{mizuki09}, 2009 & yes & 1 & $S^2_{\{0,2\}}$ \\ \hline
		$S^2_{\{2\}}$ & AND & Mizuki et al. \cite{mizuki12}, 2012 & no & 2 & $S^2_{\{0\}}$, $S^2_{\{0,1\}}$, $S^2_{\{1,2\}}$ \\ \hline
	\end{tabular}
	\medskip
	\caption{Four-card protocols to compute symmetric two-variable functions} \label{table2}
\end{table}

For $n=2$, all eight functions can be classified into three classes, as shown in Table \ref{table2}. A four-card XOR protocol was developed by Mizuki and Sone \cite{mizuki09} in 2009, and a four-card AND protocol was developed by Mizuki et al. \cite{mizuki12} in 2012.

A protocol is called \textit{committed-format} if the output is encoded in the same format as the inputs. While committed-format is a desirable property (so that the output can be used as the input in other protocols), it has been proved \cite{koch} that there is no four-card committed-format AND protocol with a guaranteed finite number of shuffles\footnote{The number of shuffles is defined to be the number of times we perform shuffling operations described in Section \ref{shuf}}, so the protocol of Mizuki et al. \cite{mizuki12} is the optimal one in this sense.

\begin{table}
	\centering
	\begin{tabular}{|c|c|c|c|c|c|}
		\hline
		\textbf{Function} & \textbf{Name} & \textbf{Protocol} & \textbf{\thead{Com-\\mitted?}} & \textbf{\thead{\#Shuf-\\fles}} & \textbf{\thead{Other Functions\\in the Same Class}} \\ \hline
		$S^3_{\emptyset}$ & Constant & \multicolumn{3}{c|}{trivial} & $S^3_{\{0,1,2,3\}}$ \\ \hline
		$S^3_{\{1,3\}}$ & XOR & Mizuki-Sone \cite{mizuki09}, 2009 & yes & 2 & $S^3_{\{0,2\}}$ \\ \hline
		\multirow{2}{*}{$S^3_{\{3\}}$} & \multirow{2}{*}{AND} & Mizuki \cite{mizuki16}, 2016 & no & 5 & \multirow{2}{*}{$S^3_{\{0\}}$, $S^3_{\{0,1,2\}}$, $S^3_{\{1,2,3\}}$} \\ \cline{3-5}
		& & Isuzugawa et al. \cite{isuzugawa}, 2021 & no & 2 & \\ \hline
		\multirow{2}{*}{$S^3_{\{0,3\}}$} & \multirow{2}{*}{Equality} & Shinagawa-Mizuki \cite{shinagawa}, 2019 & no & 1 & \multirow{2}{*}{$S^3_{\{1,2\}}$} \\ \cline{3-5}
		& & Ruangwises-Itoh \cite{ruangwises}, 2021 & yes & 2 & \\ \hline
		$S^3_{\{2,3\}}$ & Majority & Toyoda et al. \cite{toyoda}, 2021 & no & 2 & $S^3_{\{0,1\}}$ \\ \hline
		$S^3_{\{1\}}$ & 1Mod3 & Shikata et al. \cite{shikata2}, 2023 & no & 3 & $S^3_{\{2\}}$, $S^3_{\{0,1,3\}}$, $S^3_{\{0,2,3\}}$ \\ \hline
	\end{tabular}
	\medskip
	\caption{Six-card protocols to compute symmetric three-variable functions} \label{table3}
\end{table}

\subsection{Three Variables}
For $n=3$, all 16 functions can be classified into six classes, as shown in Table \ref{table3}. Several researchers independently developed protocols to compute different functions. First, as the XOR protocol of Mizuki and Sone \cite{mizuki09} is committed-format, it can be repeatedly applied to any number of variables without requiring helping cards. In 2016, Mizuki \cite{mizuki16} developed a card-minimal protocol to compute the AND function with any number of variables. Isuzugawa et al. \cite{isuzugawa} later improved the performance of the three-variable AND protocol to use only two shuffles.

In 2019, Shinagawa and Mizuki \cite{shinagawa} constructed a six-card protocol to compute the equality function $S^3_{\{0,3\}}$ that uses only a single shuffle. Later, Ruangwises and Itoh \cite{ruangwises} developed a card-minimal protocol for the same function (with any number of variables) that uses more shuffles but is committed-format. They also introduced a general technique to compute any $n$-variable \textit{doubly symmetric}\footnote{A doubly symmetric function is a function $S_X^n$ such that $x \in X$ if and only if $n-x \in X$ for every $x \in \{0,1,...,n\}$.} Boolean function with $2n$ cards. In 2021, Toyoda et al. \cite{toyoda} proposed a six-card protocol to compute the majority function $S^3_{\{2,3\}}$.

Very recently, in 2023, Shikata et al. \cite{shikata2} introduced a six-card protocol to compute any three-variable function $S_X^n$ such that $0 \in X$ if and only if $3 \in X$. This solved the only remaining non-trivial class (the one containing $S^3_{\{1\}}$) in the case $n=3$, finally making the problem settled for this case.

\subsection{Four Variables}
\begin{table}
	\centering
	\begin{tabular}{|c|c|c|c|c|c|}
		\hline
		\textbf{Function} & \textbf{Name} & \textbf{Protocol} & \textbf{\thead{Com-\\mitted?}} & \textbf{\thead{\#Shuf-\\fles}} & \textbf{\thead{Other Functions\\in the Same Class}} \\ \hline
		$S^4_{\emptyset}$ & Constant & \multicolumn{3}{c|}{trivial} & $S^4_{\{0,1,2,3,4\}}$ \\ \hline
		$S^4_{\{1,3\}}$ & XOR & Mizuki-Sone \cite{mizuki09}, 2009 & yes & 3 & $S^4_{\{0,2,4\}}$ \\ \hline
		$S^4_{\{4\}}$ & AND & Mizuki \cite{mizuki16}, 2016 & no & 5 & $S^4_{\{0\}}$, $S^4_{\{0,1,2,3\}}$, $S^4_{\{1,2,3,4\}}$ \\ \hline
		$S^4_{\{0,4\}}$ & Equality & \multirow{2}{*}{Ruangwises-Itoh \cite{ruangwises}, 2021} & yes & 3 & $S^4_{\{1,2,3\}}$ \\ \cline{1-2} \cline{4-6}
		$S^4_{\{2\}}$ & 2Mod3 & & no & 3 & $S^4_{\{0,1,3,4\}}$ \\ \hline
		$S^4_{\{1\}}$ & -- & \multirow{2}{*}{Shikata et al. \cite{shikata2}, 2023} & no & $\approx$ 7* & $S^4_{\{3\}}$, $S^4_{\{0,1,2,4\}}$, $S^4_{\{0,2,3,4\}}$ \\ \cline{1-2} \cline{4-6}
		$S^4_{\{1,2\}}$ & -- & & no & $\approx$ 8* & $S^4_{\{2,3\}}$, $S^4_{\{0,1,4\}}$, $S^4_{\{0,3,4\}}$ \\ \hline
		$S^4_{\{0,3\}}$ & 0Mod3 & \textbf{Ours (\S \ref{our})} & no & 4 & $S^4_{\{1,4\}}$, $S^4_{\{0,2,3\}}$, $S^4_{\{1,2,4\}}$ \\ \hline
		$S^4_{\{3,4\}}$ & Majority & \multicolumn{3}{c|}{\multirow{2}{*}{open problem}} & $S^4_{\{0,1\}}$, $S^4_{\{0,1,2\}}$, $S^4_{\{2,3,4\}}$ \\ \cline{1-2} \cline{6-6}
		$S^4_{\{0,2\}}$ & -- & \multicolumn{3}{c|}{} & $S^4_{\{2,4\}}$, $S^4_{\{0,1,3\}}$, $S^4_{\{1,3,4\}}$ \\ \hline
	\end{tabular}
	\medskip
	\caption{Eight-card protocols to compute symmetric four-variable functions (An asterisk denotes the expected number of shuffles in Las Vegas protocols.)} \label{table4}
\end{table}

For $n=4$, all 32 functions can be classified into ten classes, as shown in Table \ref{table4}. The aforementioned XOR protocol \cite{mizuki09}, AND protocol \cite{mizuki16}, and equality protocol \cite{ruangwises} can compute respective functions with eight cards. Also, function $S^3_{\{2\}}$ is doubly symmetric and thus can be computed by the technique of Ruangwises and Itoh \cite{ruangwises}.

Functions $S^3_{\{1\}}$ and $S^3_{\{1,2\}}$ can be computed with eight cards by protocols of Shikata et al. \cite{shikata2}. However, their protocols are Las Vegas\footnote{A Las Vegas protocol is a protocol that does not guarantee a finite number of shuffles, but has a finite expected number of shuffles.}.

The remaining three classes still lack a card-minimal protocol, leaving open problems of whether there exist such protocols. In Section \ref{our}, we propose an eight-card protocol to compute the 0Mod3 function $S^3_{\{0,3\}}$, solving one of the three open problems.

\subsection{More than Four Variables}
In 2022, Shikata et al. \cite{shikata} proved that there exists a $2n$-card protocol to compute any $n$-variable symmetric Boolean function for $n \geq 8$. This limits the open problems to only $n=5$, 6, and 7, where there are 64, 128, and 256 functions that can be classified into 20, 36, and 72 classes, respectively. 

For $n=5$, five non-trivial classes of functions can be computed by existing protocols. Our protocol in Section \ref{our} can compute the 0Mod3 function $S^5_{\{0,3\}}$, leaving 13 non-trivial classes as open problems.

For $n=6$, eight non-trivial classes of functions can be computed by existing protocols. Our protocol in Section \ref{our} can compute the 1Mod3 function $S^6_{\{1,4\}}$, leaving 26 non-trivial classes as open problems.

For $n=7$, 12 non-trivial classes of functions can be computed by existing protocols. Our protocol in Section \ref{our} can compute the 0Mod3 function $S^7_{\{0,3,6\}}$, leaving 58 non-trivial classes as open problems.

\section{Our Protocol for $k$Mod3 Function} \label{our}
For $k \in \{0,1,2\}$, the $k$Mod3 function determines whether the sum of all inputs is congruent to $k$ modulo 3. Formally, it is defined by
$$k\text{Mod3}(x_1,x_2,...,x_n) = \begin{cases}
	1, &\text{ if } \sum_{i=1}^n x_i \equiv k\text{ (mod 3);} \\
  0, &\text{ otherwise.}
\end{cases}$$

In this section, we will briefly describe the necessary subprotocols, then we will introduce our $k$Mod3 protocol that uses $2n$ cards for any $n \geq 3$.

\subsection{Preliminaries} \label{shuf}
We use two types of cards in our protocol: \mybox{$\clubsuit$} and \mybox{$\heartsuit$}. All cards have indistinguishable back sides. As per convention, we encoded 0 by a commitment \hbox{\mybox{$\clubsuit$}\mybox{$\heartsuit$}}, and 1 by a commitment \hbox{\mybox{$\heartsuit$}\mybox{$\clubsuit$}}.

All of the randomness in our protocol is generated by the following two shuffling operations, which are jointly executed by all parties. (For other operations that do not involve randomness, the operations can be executed by any party while being observed by other parties.)

\subsubsection{Random Cut}
Given a sequence $S$ of $n$ cards, a \textit{random cut} shifts $S$ by a uniformly random cyclic shift unknown to all parties. It can be implemented by letting all parties take turns to apply \textit{Hindu cuts} (taking several card from the bottom of the pile and putting them on the top) to $S$.

\subsubsection{Random $k$-Section Cut}
Given a sequence $S$ of $kn$ cards, a \textit{random $k$-section cut} \cite{polygon} divides $S$ into $k$ blocks, each consisting of $n$ consecutive cards, then shifts the blocks by a uniformly random cyclic shift unknown to all parties. It can be implemented by putting all cards in each block into an envelopes and applying the random cut to the sequence of envelopes.

\subsection{Encoding Integers in $\mathbb{Z}/3\mathbb{Z}$}
For $i \in \{0,1,2\}$, define $E_3^\clubsuit(i)$ to be a sequence of three cards, all of them being \hbox{\mybox{$\heartsuit$}s} except the ($i+1$)-th card from the left being a \mybox{$\clubsuit$}, e.g. $E_3^\clubsuit(1)$ is \hbox{\mybox{$\heartsuit$}\mybox{$\clubsuit$}\mybox{$\heartsuit$}}. Conversely, define $E_3^\heartsuit(i)$ to be a sequence of three cards, all of them being \hbox{\mybox{$\clubsuit$}s} except the ($i+1$)-th card from the left being a \mybox{$\heartsuit$}, e.g. $E_3^\heartsuit(0)$ is \hbox{\mybox{$\heartsuit$}\mybox{$\clubsuit$}\mybox{$\clubsuit$}}.

\subsection{Adding Two Commitments in $\mathbb{Z}/3\mathbb{Z}$} \label{add1}
Given two commitments of bits $a,b \in \{0,1\}$, this subprotocol produces either $E_3^{\clubsuit}(a+b)$ or $E_3^{\heartsuit}(a+b)$, each with probability 1/2, without using any helping card. It was developed by Shikata et al. \cite{shikata}.

\begin{enumerate}
	\item Arrange the commitments of $a$ and $b$ as a sequence in this order from left to right.
	\item Apply a random 2-section cut on the sequence. Then, apply a random cut on the middle two cards.
	\item Let $(c_1,c_2,c_3,c_4)$ be the obtained sequence in this order from left to right. Turn over $c_2$.
	\begin{itemize}
		\item If it is a \mybox{$\clubsuit$}, then $(c_1,c_3,c_4)$ will be $E_3^{\clubsuit}(a+b)$.
		\item If it is a \mybox{$\heartsuit$}, then $(c_4,c_3,c_1)$ will be $E_3^{\heartsuit}(a+b)$. 
	\end{itemize}
\end{enumerate}

\subsection{Adding Two Integers in $\mathbb{Z}/3\mathbb{Z}$} \label{add2}
Given two sequences $E_3^{\clubsuit}(a)=(a_0,a_1,a_2)$ and $E_3^{\heartsuit}(b)=(b_0,b_1,b_2)$, this subprotocol produces a sequence $E_3^{\clubsuit}(a+b\text{ mod }3)$ without using any helping card. It was developed by Ruangwises and Itoh \cite{ruangwises}.

\begin{enumerate}
	\item Rearrange the cards as a sequence $(a_0,b_2,a_1,b_1,a_2,b_0)$.
	\item Apply a random 3-section cut on the sequence, transforming it into $(a_r,b_{2-r},a_{r+1},$ $b_{1-r},a_{r+2},b_{-r})$ for a uniformly random $r \in \mathbb{Z}/3\mathbb{Z}$, where the indices are taken modulo $3$.
	\item Rearrange the cards back as they were before. We now have two sequences $E_3^{\clubsuit}(a-r\text{ mod }3)=(a_r,a_{r+1},a_{r+2})$ and $E_3^{\heartsuit}(b+r\text{ mod }3)=(b_{-r},b_{1-r},b_{2-r})$, where the indices are taken modulo $3$.
	\item Turn over the sequence $E_3^{\heartsuit}(b+r\text{ mod }3)$ to reveal $s = b+r\text{ mod }3$. Then, shift the sequence $E_3^{\clubsuit}(a-r)$ cyclically to the right by $s$ positions, transforming it into $E_3^{\clubsuit}(a-r+s\text{ mod }3)=E_3^{\clubsuit}(a+b\text{ mod }3)$ as desired.
\end{enumerate}

\subsection{Main Protocol}
We use an idea similar to the one in \cite{shikata2}, but extends their idea further to the case of $n$ variable for any $n \geq 3$.
\begin{enumerate}
	\item Apply the subprotocol in Section \ref{add1} on the commitments of $x_1$ and $x_2$ to obtain either $E_3^{\clubsuit}(x_1+x_2)$ or $E_3^{\heartsuit}(x_1+x_2)$.
	\item If we get an $E_3^{\clubsuit}(x_1+x_2)$, use a free \mybox{$\clubsuit$} and the commitment of $x_3$ to create $E_3^{\heartsuit}(x_3)$. Then, apply the subprotocol in Section \ref{add2} on $E_3^{\clubsuit}(x_1+x_2)$ and $E_3^{\heartsuit}(x_3)$ to obtain $E_3^{\clubsuit}(x_1+x_2+x_3\text{ mod }3)$. Conversely, if we get an $E_3^{\heartsuit}(x_1+x_2)$, use a free \mybox{$\heartsuit$} and the commitment of $x_3$ to create $E_3^{\clubsuit}(x_3)$. Then, apply the subprotocol in Section \ref{add2} on $E_3^{\clubsuit}(x_3)$ and $E_3^{\heartsuit}(x_1+x_2)$ to obtain $E_3^{\clubsuit}(x_1+x_2+x_3\text{ mod }3)$.
	\item Use a free \mybox{$\clubsuit$} and the commitment of $x_4$ to create $E_3^{\heartsuit}(x_4)$. Then, apply the subprotocol in Section \ref{add2} on $E_3^{\clubsuit}(x_1+x_2+x_3)$ and $E_3^{\heartsuit}(x_4)$ to obtain $E_3^{\clubsuit}(x_1+x_2+x_3+x_4\text{ mod }3)$. Repeatedly perform this step for the rest of inputs until we obtain $E_3^{\clubsuit}(x_1+x_2+...+x_n\text{ mod }3)$.
	\item Turn over the $(k+1)$-th leftmost card. If it is a \mybox{$\clubsuit$}, return 1; otherwise, return 0.
\end{enumerate}

\subsection{Proof of Correctness and Security}
Our main protocol largely depends on the subprotocols in Sections \ref{add1} and \ref{add2}. Proofs of correctness and security of these two subprotocols are shown in \cite[\S 3.1]{shikata} and \cite[\S A.1]{ruangwises}, respectively. The proofs consist of drawing a structure called \textit{KWH-tree}, which iterates all possible states and their probabilities of the sequence of cards after each operation. By observing the conditional probability of each state, one can verify that (1) the resulting sequence is always correct, and (2) turning some cards face-up during the protocol does not reveal any probabilistic information of the inputs.

Assuming the correctness and security of both subprotocols, it is easy to show that our main protocol is also correct and secure.

In Step 1, the probability of getting $E_3^{\clubsuit}(x_1+x_2)$ or $E_3^{\heartsuit}(x_1+x_2)$ is 1/2 each, regardless of the inputs \cite[\S 3.1]{shikata}. Therefore, one cannot deduct any information of $x_1$ and $x_2$ upon getting $E_3^{\clubsuit}(x_1+x_2)$ or $E_3^{\heartsuit}(x_1+x_2)$.

Steps 2 and 3 are straightforward applications of the subprotocol in Section \ref{add2}, and thus are correct and secure.

In Step 4, the resuting sequence is $E_3^{\clubsuit}(x_1+x_2+...+x_n\text{ mod }3)$, which is either \hbox{\mybox{$\clubsuit$}\mybox{$\heartsuit$}\mybox{$\heartsuit$}}, \hbox{\mybox{$\heartsuit$}\mybox{$\clubsuit$}\mybox{$\heartsuit$}}, or \hbox{\mybox{$\heartsuit$}\mybox{$\heartsuit$}\mybox{$\clubsuit$}}. Turning over the $(k+1)$-th leftmost card reveals whether $x_1+x_2+...+x_n \equiv k\text{ (mod 3)}$ (if the card is a \mybox{$\clubsuit$}, then $x_1+x_2+...+x_n \equiv k\text{ (mod 3)}$; if the card is a \mybox{$\heartsuit$}, then $x_1+x_2+...+x_n \not\equiv k\text{ (mod 3)}$) without revealing any other information of the inputs. Hence, our main protocol is correct and secure.

\subsection{Analysis}
\begin{table}
	\centering
	\begin{tabular}{|c|c|c|c|}
		\hline
		\textbf{Protocol} & \textbf{\#Cards} & \textbf{\#Shuffles} & \textbf{Committed?} \\ \hline
		Nishida et al. \cite{nishida}, 2015 & $2n+2$ & $O(n \lg n)$ & yes \\ \hline
		Ruangwises-Itoh \cite{ruangwises}, 2021 & $2n+2$ & $n$ & no \\ \hline
		\textbf{Ours} & $2n$ & $n$ & no \\ \hline
	\end{tabular}
	\medskip
	\caption{Properties of protocols to compute the $k$Mod3 function} \label{table5}
\end{table}

Our protocol is the first card-minimal protocol for the $k$Mod3 function. There are several existing protocols that can compute the function using more than $2n$ cards, e.g. the protocol of Nishida et al. \cite{nishida} which uses $2n+2$ cards and $O(n \lg n)$ shuffles, and the protocol of Ruangwises-Itoh \cite{ruangwises} which uses $2n+2$ cards and $n$ shuffles. See Table \ref{table5}. Therefore, our protocol is also optimal in terms of number of shuffles.

For each $n \geq 3$, our protocol can compute six functions (0Mod3, 1Mod3, 2Mod3, and their negations), which are from two different NPN-equivalence classes (for $n \equiv 0\text{ (mod 3)}$, 1Mod3 and 2Mod3 are in the same class; for $n \equiv 1\text{ (mod 3)}$, 0Mod3 and 1Mod3 are in the same class; for $n \equiv 2\text{ (mod 3)}$, 0Mod3 and 2Mod3 are in the same class).

\section{Future Work}
We formulated the problem of developing $2n$-card protocols to compute $n$-variable symmetric Boolean functions, and also proposed protocols for some classes of these functions. It remains an open problem to construct card-minimal protocols, or to prove that none exists, for the remaining classes in the cases $n=4$, 5, 6, and 7.

Another possible consideration is the property of a protocol. For example, if we restrict the protocols to be the ones with a guaranteed finite number of shuffles, functions $S^4_{\{1\}}$ and $S^4_{\{1,2\}}$ still lack a card-minimal finite protocol, leaving four classes unsolved in the case $n=4$. Also, most of the existing protocols are not committed-format. It is a challenging work to construct card-minimal committed-format protocols to compute more functions, or to prove that none exists for some functions, similar to the proof of non-existence of a four-card committed-format finite AND protocol in \cite{koch}.

\subsubsection*{Acknowledgement}
The author would like to thank Daiki Miyahara for a valuable discussion on this research.


\begin{thebibliography}{}
	\bibitem{denboer} B. den Boer. More Efficient Match-Making and Satisfiability: the Five Card Trick. In \textit{Proceedings of the Workshop on the Theory and Application of of Cryptographic Techniques (EUROCRYPT '89)}, pp. 208--217 (1990).
	\bibitem{isuzugawa} R. Isuzugawa, K. Toyoda, Y. Sasaki, D. Miyahara and T. Mizuki. A Card-Minimal Three-Input AND Protocol Using Two Shuffles. In \textit{Proceedings of the 27th International Computing and Combinatorics Conference (COCOON)}, pp. 668--679 (2021).
	\bibitem{landscape} A. Koch. The Landscape of Optimal Card-Based Protocols. \textit{Mathematical Cryptology}, 1(2): 115--131 (2021).
	\bibitem{koch} A. Koch, S. Walzer and K. H\"{a}rtel. Card-Based Crypto-graphic Protocols Using a Minimal Number of Cards. In \textit{Proceedings of the 21st International Conference on the Theory and Application of Cryptology and Information Security (ASIACRYPT)}, pp. 783--807 (2015).
	\bibitem{mizuki16} T. Mizuki. Card-based protocols for securely computing the conjunction of multiple variables. \textit{Theoretical Computer Science}, 622: 34--44 (2016).
	\bibitem{mizuki12} T. Mizuki, M. Kumamoto and H. Sone. The Five-Card Trick Can Be Done with Four Cards. In \textit{Proceedings of the 18th International Conference on the Theory and Application of Cryptology and Information Security (ASIACRYPT)}, pp. 598--606 (2012).
	\bibitem{mizuki09} T. Mizuki and H. Sone. Six-Card Secure AND and Four-Card Secure XOR. In \textit{Proceedings of the 3rd International Frontiers of Algorithmics Workshop (FAW)}, pp. 358--369 (2009).
	\bibitem{nishida} T. Nishida, Y. Hayashi, T. Mizuki and H. Sone. Card-Based Protocols for Any Boolean Function. In \textit{Proceedings of the 12th Annual Conference on Theory and Applications of Models of Computation (TAMC)}, pp. 110--121 (2015).
	\bibitem{ruangwises} S. Ruangwises and T. Itoh. Securely Computing the $n$-Variable Equality Function with $2n$ Cards. \textit{Theoretical Computer Science}, 887: 99--100 (2021).
	\bibitem{class} T. Sasao. Switching Theory for Logic Synthesis, 1st edn. Kluwer Academic Publishers, Norwell (1999).
	\bibitem{shikata2} H. Shikata, D. Miyahara and T. Mizuki. Few-helping-card Protocols for Some Wider Class of Symmetric Boolean Functions with Arbitrary Ranges. In \textit{Proceedings of the 10th ACM International Workshop on ASIA Public-Key Cryptography (APKC)}, pp. 33--41 (2023).
	\bibitem{shikata} H. Shikata, D. Miyahara, K. Toyoda and T. Mizuki. Card-Minimal Protocols for Symmetric Boolean Functions of More than Seven Inputs. In \textit{Proceedings of the 19th International Colloquium on Theoretical Aspects of Computing (ICTAC)}, pp. 388--406 (2022).
	\bibitem{shinagawa} K. Shinagawa and T. Mizuki. The Six-Card Trick: Secure Computation of Three-Input Equality. In \textit{Proceedings of the 21st Annual International Conference on Information Security and Cryptology (ICISC)}, pp. 123--131 (2018).
	\bibitem{polygon} K. Shinagawa, T. Mizuki, J.C.N. Schuldt, K. Nuida, N. Kanayama, T. Nishide, G. Hanaoka and E. Okamoto. Card-Based Protocols Using Regular Polygon Cards. \textit{IEICE Transactions on Fundamentals of Electronics, Communications and Computer Sciences}, E100.A(9): 1900--1909 (2017).
	\bibitem{toyoda} K. Toyoda, D. Miyahara and T. Mizuki. Another Use of the Five-Card Trick: Card-Minimal Secure Three-Input Majority Function Evaluation. In \textit{Proceedings of the 22nd International Conference on Cryptology in India (INDOCRYPT)}, pp. 536--555 (2021).
\end{thebibliography}
\end{document}